# Breaking the doping limit in silicon by deep impurities

Mao Wang[1,2,*], A. Debernardi[3,**], Y. Berencén[1], R. Heller[1], Chi Xu[1,2], Ye Yuan[1,4], Yufang Xie[1,2],

R. Böttger[1], L. Rebohle[1], W. Skorupa[1], M. Helm[1,2], S. Prucnal[1] and Shengqiang Zhou[1,*]

[1]Helmholtz-Zentrum Dresden-Rossendorf, Institute of Ion Beam Physics and Materials Research, Bautzner Landstr. 400, 01328 Dresden, Germany
[2]Technische Universität Dresden, 01062 Dresden, Germany
[3]CNR-IMM, sede Agrate Brianza, via Olivetti 2, I-20864, Agrate Brianza, Italy
[4]Physical Science and Engineering Division (PSE), King Abdullah, University of Science and Technology (KAUST), Thuwal 23955-6900, Saudi Arabia

## Abstract

N-type doping in Si by shallow impurities, such as P, As and Sb, exhibits an intrinsic limit due to the Fermi-level pinning via defect complexes at high doping concentrations. Here we demonstrate that doping Si with the chalcogen Te by non-equilibrium processing, a deep double donor, can exceed this limit and yield higher electron concentrations. In contrast to shallow impurities, both the interstitial Te fraction decreases with increasing doping concentration and substitutional Te dimers become the dominant configuration as effective donors, leading to a non-saturating carrier concentration as well as to an insulator-to-metal transition. First-principle calculations reveal that the Te dimers possess the lowest formation energy and donate two electrons per dimer to the conduction band. These results provide novel insight into physics of deep impurities and lead to a possible solution for the ultra-high electron concentration needed in today's Si-based nanoelectronics.

[*]Corresponding authors, email: m.wang@hzdr.de; s.zhou@hzdr.de

[**]Corresponding author [theory], e-mail: alberto.debernardi@mdm.imm.cnr.it

Scaling evolution for the future generations of Si-based technology requires extremely high free-electron densities (around $10^{21}$ cm$^{-3}$) [1,2]. This is usually achieved with shallow dopants (e.g. group V dopants such as P, As and Sb) due to their low ionization energies (20-50 meV), low diffusivities, and suitable solid solubilities in Si. Current technologies, implantation or molecular beam epitaxy, can achieve doping concentrations greater than the solid solubility limit. However, the free-electron concentrations are still found to saturate at around $5\times10^{20}$ cm$^{-3}$ [3-5]. The electrical activation approaches a limit where the additional introduced dopant atoms cease to generate extra free carriers. During the past decades, experimental results [6-8] and first-principles calculations [9-11] were applied to investigate the microscopic nature of the electrical deactivation of donors in Si. The As$_n$$V$ (n ≤ 4) model (clustering around a vacancy surrounded by As atoms) [10-17] and the dimer model (two donor atoms bound to reconfigured Si with no vacancies) [18,19] are two classical models for electrically deactivating dopant complexes. The latter, however, has been suggested to be dominant at high doping levels [19,20]. The donor dimers introduce localized deep-level states in the band gap and compensate the free electrons, preventing high electron concentrations [18,19].

Alternatively, it has been found that deep-level impurities (e.g. chalcogen dopants) with ionization energies of several hundred meV can also induce free electrons (in excess of $10^{20}$ cm$^{-3}$) in Si accompanied with an insulator-to-metal transition (IMT) [21,22]. This makes chalcogens good prospects to be used for overcoming the doping limit in Si. Particularly, one puzzling observation in chalcogen-doped Si is that the electron concentration increases linearly with the dopant concentration while the substitutional fraction remains almost constant [23,24]. For traditional shallow-level dopants, interstitials, inactive clusters and precipitates often become more energetically favorable as the dopant concentration increases, e.g. in As-doped Si [25,26], Mn-doped GaAs [27,28] and Mg-doped GaN [29,30]. Therefore, scrutinizing the atomic configuration of chalcogen impurities in Si is needed to understand this anomalous electrical activation behavior as well as the sub-bandgap infrared photoresponse [31-33].

In this Letter, we unveil the microscopic origin for the high electron density that can be achieved by chalcogen doping. We use Rutherford backscattering channeling to directly identify the lattice location of Te in Si and correlate the Te atomistic configuration with the corresponding electrical properties. It is found that at higher Te concentration the fraction of substitutional Te increases, contrary to the behavior of shallow dopants. By first-principles



calculations, we find that substitutional Te dimers are energetically more favorable and drive the occurrence of non-saturating free-electron densities leading to an insulator-to-metal transition. We thus propose chalcogen dimers as effective donors in Si for ultra-high *n*-type doping.

Ion implantation into an intrinsic Si substrate combined with pulsed laser melting (PLM) was applied to synthesize single-crystalline and epitaxial Te-hyperdoped Si layers with Te concentration several orders of magnitude above the solid solubility limit. The samples were labelled as Te-0.25%...Te-2.5% with the numbers standing for the percentage of the Te peak concentration. Rutherford backscattering spectrometry/channeling (RBS/C) measurements were performed using a 1.7 MeV $He^+$ beam. The backscattered particles were detected at an angle of 170° with respect to the incoming beam direction using silicon surface barrier detectors. Angular axial scans around different crystallographic directions were obtained using a two-axis goniometer. The electrical properties of the PLM-treated Te-hyperdoped layers were measured using a commercial Lake Shore Hall Measurement System in a van der Pauw configuration [34]. The structural and electronic calculations in the framework of the Density Functional Theory (DFT) were performed by solving the Kohn-Sham equations through the plane-wave pseudopotential approach implemented in the Quantum Espresso (QE) open-source code [35]. Further details about experiments and computations can be found in the supplementary information.

Figure 1(a) shows the carrier concentration of Te-hyperdoped Si samples measured by Hall effect at 300 K as a function of doping concentration. The carrier concentration shown here is in the range of $2.0 \times 10^{19}$ $cm^{-3}$ to $8.3 \times 10^{20}$ $cm^{-3}$. Being completely different from previously published results for As, P and Sb doped Si [19-20, 36-43], the electron concentration increases monotonically as the doping concentration increases and does not show any sign of saturation even at an atomic dopant concentration as high as 2.5% (exceeding $10^{21}$ $cm^{-3}$). The unsaturated carrier concentration in the Te-hyperdoped Si layers is approaching $10^{21}$ $cm^{-3}$, which is the value required for the next generation of Si technology [1,4,44]. This shows an outstanding behavior of deep impurities as compared with traditional *n*-type dopants in Si (e.g. As and Sb) at very high doping levels, where the reported carrier concentration is never found to be larger than $6.5 \times 10^{20}$ $cm^{-3}$ [45-47]. In addition, FIG. 1(b) shows the room-temperature resistivity of Te-hyperdoped layers as the function of dopant concentration. Previous results for group V dopants (i.e. As, P) doped Si [36,38-40,48] are also included as references. The smallest room-temperature resistivity of Te-hyperdoped Si



samples is as low as $7.5 \times 10^{-4}$ $\Omega$cm and is comparable to the smallest resistivity value for As-doped Si. IMT is also observed in Te-hyperdoped samples as shown by temperature-dependent conductivity (FIG. S1 in the supplement). Samples Te-0.25% and Te-0.50% show an insulating behavior with a strong temperature-dependent conductivity. However, samples with the Te peak concentration exceeding 1.0% present a radically different behavior, exhibiting a conductivity with negligible variation with temperature and a finite value of more than $10^2$ (S/cm) as the temperature tends to zero. It is therefore crucial to understand this anomalous dependence of the electrical activation on the dopant concentration in Te-hyperdoped Si by scrutinizing the dopant lattice configuration at the atomic scale.

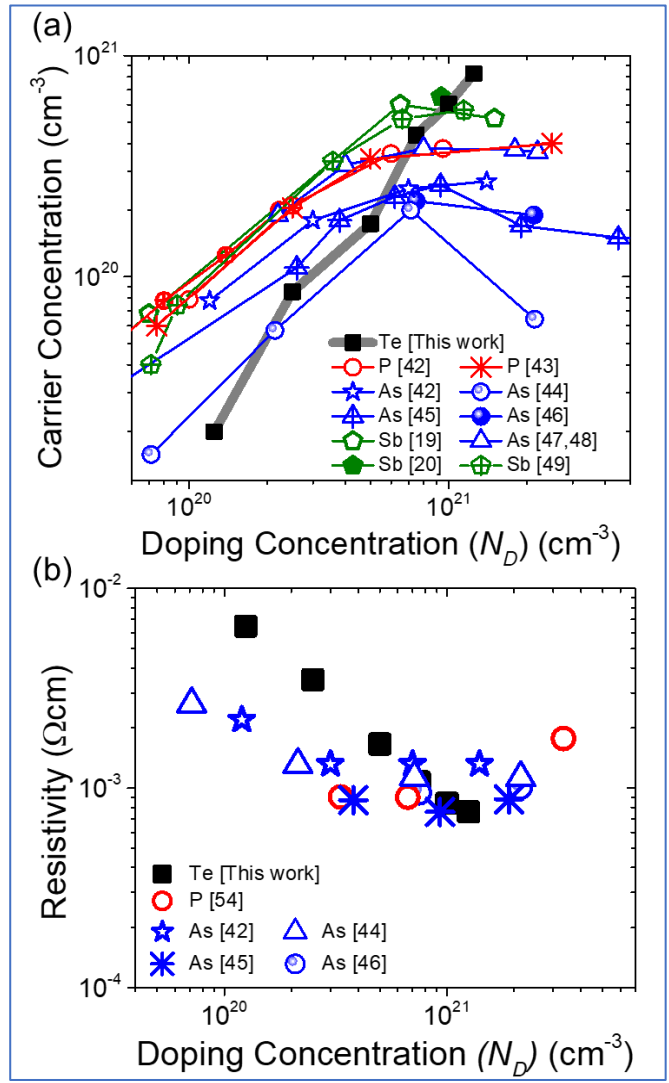

FIG. 1. (a) The carrier concentration of Te-hyperdoped Si samples measured by Hall effect at 300 K as a function of doping concentration. The carrier concentration approaches $10^{21}$ cm$^{-3}$. Previously published results for As, P and Sb heavily doped Si are also included as references. Our results overcome the limit of free-electron concentration ($5\sim6.5 \times 10^{20}$ cm$^{-3}$) obtained by the conventional *n*-type dopants in Si. Saturation in terms of the



doping level is not observed within the range of Te concentrations which are several orders of magnitude above the solid solubility limit of Te in Si. (b) The resistivity of PLM-treated Te-hyperdoped Si samples as a function of doping concentration measured at 300 K. Previous published works for As, P and Sb heavily doped Si are also included as reference.

RBS/C is a unique technique to measure the lattice location of the dopants [49] (see the supplementary information for further details). FIG. 2 shows the angular distributions of He ions backscattered by Te and Si as a function of the tilt angle around the Si <100> axes in the selected samples. More angular scans in the vicinity of the three main crystallographic directions (<100>, <110> and <111>) are displayed in the supplementary information. Two typical parameters are used to characterize the angular distributions, $\chi_{min}$ and $\psi_{1/2}$. The minimum yield in the angular scan, $\chi_{min}$, qualitatively describes the ability of dopants to block the channels. The critical angle $\psi_{1/2}$ represents the half width at half maximum of the angular scans between $\chi_{min}$ and 1. It is related to the displacement of substitutional atoms [50-52]. Both parameters have been calculated for Si and Te at the same depth. The main results and their implications are as follows:

*i)* For sample Te-0.25%, there is a peak in the middle of the angular scan. Some Te ions are located at low-symmetry interstitial sites lying around the axes of the crystal lattice. This peak decreases with increasing Te concentration. At low concentrations, Te ions occupy both substitutional and interstitial sites, whereas at higher concentration they preferably go to substitutional sites.

*ii)* The interstitial fraction estimated from the angular scan drastically decreases at Te concentrations up to 1.0% and tends to a small constant value at higher concentrations (FIG. 2(d)). It is worth noting that 1.0% is also the critical concentration where the IMT takes place in Te-doped Si (see FIG. 1(a)). This indicates a strong correlation between the atomic configuration of Te ions and the electronic properties of the samples.

*iii)* At higher concentrations, $\psi_{1/2}$-Te is narrower compared to $\psi_{1/2}$-Si. This suggests that the Te dopants do not perfectly substitute the Si lattice sites. There exists a displacement ($r_0$) from the ideal substitutional site [49]. Based on the experimental data, the projections of the displacement $r_0$ of Te atoms perpendicular to the three main axes can be estimated (see details in the supplementary information). The average value $r$ obtained from the angular scan is found to be around 0.42±0.03 Å. This is in agreement with the average displacement ($r_0 = 0.46$ Å) obtained from the first-principles calculations shown later.



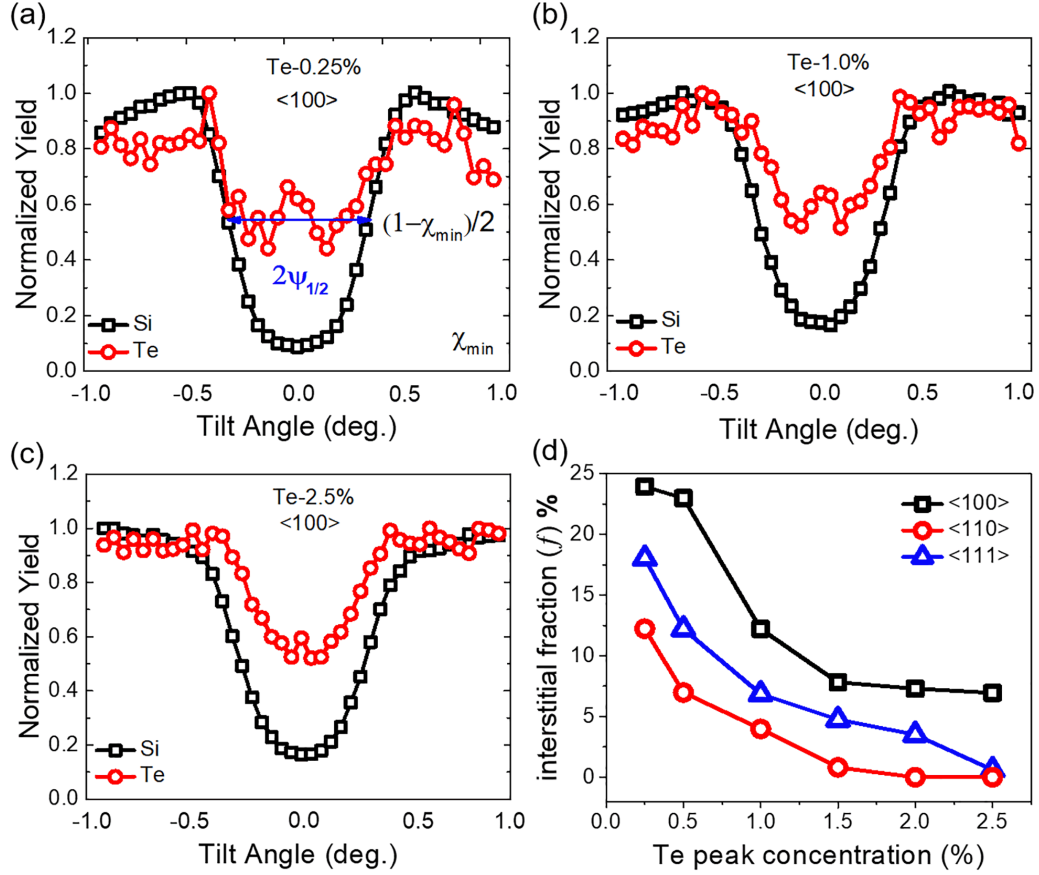

FIG. 2. (a), (b) and (c) the angular distributions about the <100> axes for 1.7 Mev He$^+$ scattered from Te (red circle) and Si (black square) for the selected samples with different Te concentrations. The angular scan of substitutional Te will overlap with that of Si as reported. A peak is expected to arise in the middle of the angular scan for interstitial Te ions. The angular scan of substitutional but slightly displaced Te atoms is expected to be narrow and shallow. (d) The interstitial fraction of Te as a function of the concentration as calculated from the angular scan results.

To further prove that the majority of Te ions are located at the substitutional lattice sites at high Te concentrations, we performed angular-scan mapping around three major crystalline axes. FIG. 3 shows representative angular-dependent two-dimensional backscattering-yield patterns of the sample with highest Te concentration (Te-2.5%). The backscattering yield patterns are extracted from the integral backscattering signals of Si and Te in the vicinity of <100>, <110> and <111> in sample Te-2.5%. Prominent channeling effects are observed for both the Si and the Te signals and their patterns show a high similarity along <100>, <110> and <111> orientations and the closest-packed planes [{100}, {110} and {111}]. These angular-scan mappings unambiguously demonstrate that the Te-hyperdoped Si layer is epitaxially regrown during the PLM treatment and that the majority of Te atoms occupy the substitutional sites of the Si matrix at such high Te concentrations.



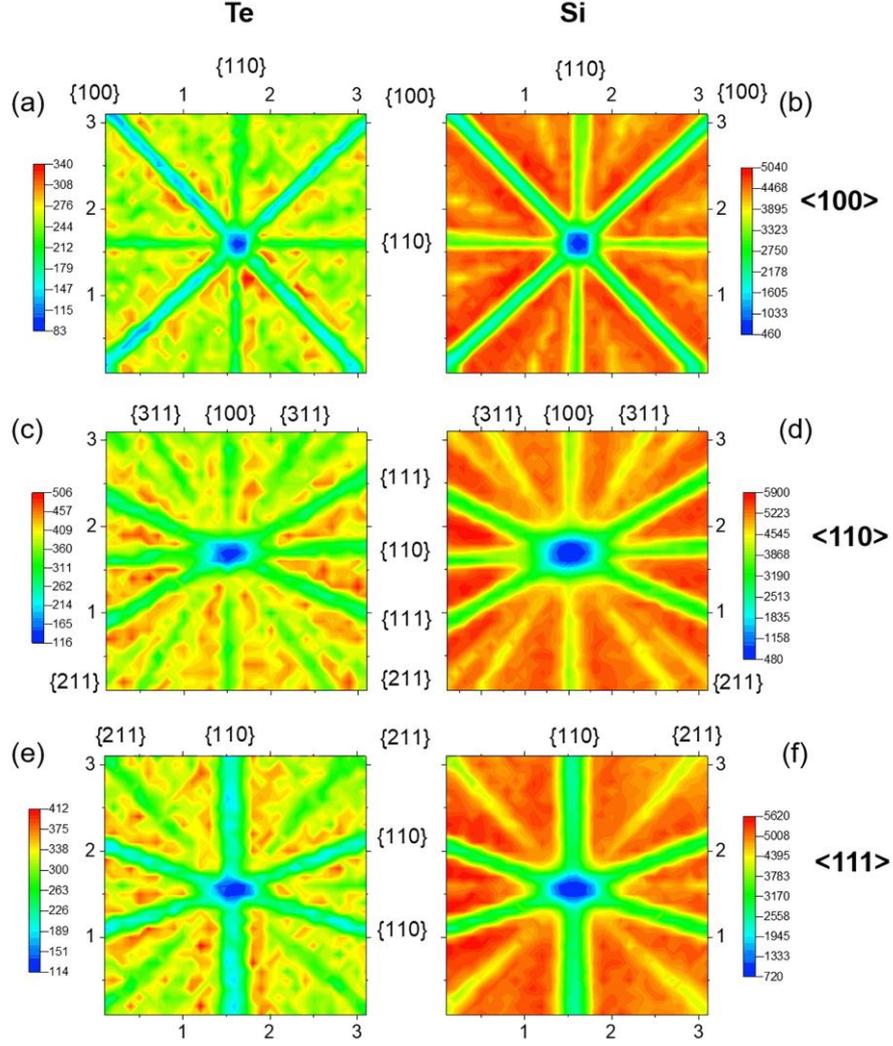

FIG. 3. Experimental two-dimensional backscattered-yield patterns of the sample Te-2.5% obtained from RBS/channeling spectra. The patterns are extracted from the integral backscattering signal from Te ((a), (c) and (e)) and Si ((b), (d) and (f)) in the vicinity of <100>, <110> and <111> directions. The patterns for the Si and the Te signals match well with each other, indicating that the majority of Te impurities are substitutional.

We directly probed the lattice location of Te ions in hyperdoped Si by RBS/Channeling. These results suggest that the decrease of Te interstitials is accompanied by an increase of displacements of Te atoms from the ideal substitutional site. To understand this, first-principles calculations (see supplementary information for further computational details) have been carried out to compute the formation energy for three types of defects as a function of Te doping concentration. We define the formation energy of a defect (D) as:

$$\Delta E_F \equiv E_D^{SC} - n_{Si}^{SC}\mu_{Si} - n_{Te}^{SC}\mu_{Te} \quad (1),$$



where $E_D^{SC}$ is the total energy of the super-cell (SC) with defect D, $n_{Si}^{SC}$ and $n_{Se}^{SC}$ are the numbers of Si and Te atoms in the super-cell, while the chemical potentials $\mu_{Si}$ and $\mu_{Te}$ correspond to pure Si and SiTe$_2$ in equilibrium with each other. In FIG. 4(a) we summarize our results for the most relevant types of defects: substitutional single Te ($S_{Te}$) and substitutional Te dimer ($S_{Te}$-$S_{Te}$). Moreover, the results for interstitial defects ($I_{Te}$ at hexagonal *H* and tetrahedral *T* symmetry sites) at the selected concentrations are displayed in the inset of FIG. 4(a) for comparison. Generally, the formation energy of the interstitial sites (*T* site and *H* site) is much larger (several eV per Te atom) than that of the substitutional site in the Te-hyperdoped Si system, which results in an energetically unfavorable formation of interstitial Te. Surprisingly, the $S_{Te}$-$S_{Te}$ dimer presents the lowest formation energy among all types of defects considered, and the difference in the formation energy (per Te atom) between a single $S_{Te}$ dopant and a $S_{Te}$-$S_{Te}$ dimer is around 0.3 eV. This suggests that $S_{Te}$-$S_{Te}$ dimers are more energetically favorable. Note that at low concentration, the average distance between Te atoms might be too large to promote the formation of dimers. As shown in FIG. 4(a), the $S_{Te}$ formation energy increases sharply with increasing Te concentration on the insulating side which is consistent with the previously reported results about Se-hyperdoped Si [22]. However, in the regime exceeding the threshold for the IMT, the defect formation energy is relatively stable, only slightly increases with the Te concentration. Therefore, as the doping Te concentration increases, the penalty for assigning additional Te atoms becomes less costly and ultimately renders the $S_{Te}$-$S_{Te}$ dimer configuration stable. If $S_{Te}$-$S_{Te}$ dimers form, the lattice positions of Te will be slightly displaced from the ideal substitutional sites. From the *ab initio* relaxed positions of Te dimers, the displacement is estimated to be around 0.46 Å for samples with a Te concentration of 1.36%. The $S_{Te}$-$S_{Te}$ dimer formation and the displacement are in excellent agreement with the findings obtained from the experimental results shown in FIG. 2 and FIG. 3.



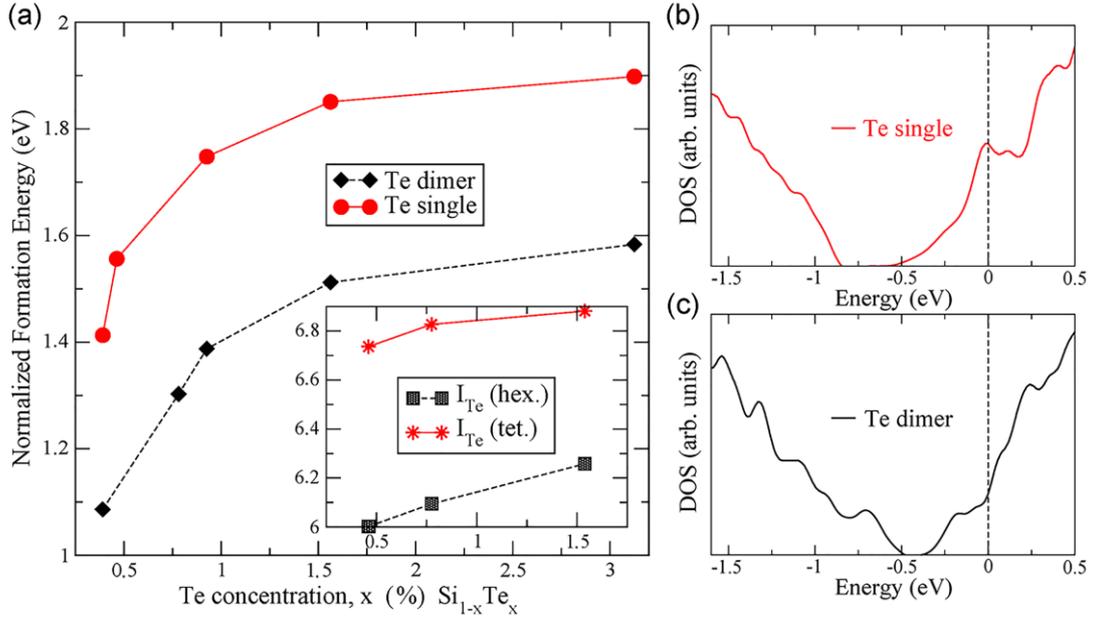

FIG. 4. (a) Normalized (per atom) formation energy ($\Delta E_F/n_{Te}^{SC}$) of isolated substitutional Te (circles) and Te-Te dimer (diamonds) as a function of Te concentration, computed via *ab inito* calculations. Inset: Formation energy of high symmetry interstitial sites (*T* and *H* denoted with squares and stars, respectively) as a function of Te concentration; lines are a guide for eyes. *Ab initio* calculations of density of states (DOS) at a Te concentration x= 1.56% for single Te dopant (b) and $S_{Te}$-$S_{Te}$ dimer (c). The zero of the energy scales (dashed lines) corresponds to the Fermi energy.

In FIG. 4(b, c) we show the density of states (DOS) corresponding to our simulations of $S_{Te}$-$S_{Te}$ and single $S_{Te}$ at a Te concentration of 1.56%, i.e. in the metallic regime. The region where the DOS is zero (a fraction of eV wide, approximately centered at -0.7 eV for $S_{Te}$ and -0.4 eV for $S_{Te}$-$S_{Te}$, respectively) separates the conduction band from the valence band, and it results from the modification of the bulk Si bandgap caused by doping. In both panels, the position of the Fermi energy denotes a metallic phase in agreement with the experimental data in FIG. 1(a). By integration of the local DOS from conduction band minimum (CBM) up to the Fermi energy we computed the number of carriers provided by each type of defects. From our *ab initio* simulations, we found that a single $S_{Te}$ acts as a double donor, which is consistent with other reports [53,54]. One $S_{Te}$-$S_{Te}$ dimer provides only two electrons to the conduction band, i.e. each Te composing the dimer provides only one electron to the conduction band. From FIG. 1 we can notice that at high Te concentration, the electron concentration ($8.1 \times 10^{20}$ cm$^{-3}$) is approaching the dopant concentration ($1.25 \times 10^{21}$ cm$^{-3}$) due to the formation of $S_{Te}$-$S_{Te}$ dimers. The lowest resistivity in our films is also comparable with As- and P-doped Si for similar doping levels (see the supplementary information for further details). We anticipate that the electron concentration can be further increased by optimizing



the Te implantation fluence and annealing parameters, therefore opening a new avenue towards ultra-high *n*-type doping for the Si-based new generation microelectronics. As a final remark, we cannot exclude the appearance of Te$_n$*V* complexes from RBS/C [55,56]. However, the measured high carrier concentration supports that Te$_n$*V* complexes cannot be the major states since they are electrically-inactive centers [7,12,14,16,20].

In conclusion, we have demonstrated that the long-standing *n*-type doping limit in Si can be overcome by using deep-level donors (e.g. Te) instead of traditional shallow-level donors (e.g. P, As, Sb). We have provided direct experimental and theoretical evidences for the formation of $S_{Te}$-$S_{Te}$ dimers with increasing Te concentration that have been found to be responsible for the non-saturating electron concentrations and the insulator-to-metal transition. This work provides an insight to understand the lattice configurations and electronic structures of deep-level dopants in semiconductors.


## ACKNOWLEDGMENTS

The authors thank Wolfhard Möller for fruitful discussion. Support by the Ion Beam Center (IBC) at HZDR is gratefully acknowledged. This work is funded by the Helmholtz-Gemeinschaft Deutscher Forschungszentren (HGF-VH-NG-713). M.W. acknowledges financial support by Chinese Scholarship Council (File No. 201506240060). A.D. acknowledges CINECA for computational resources allocated under ISCRA initiative (IMeTe project) and thanks R. Colnaghi for technical support on computer hardware. Y.B. would like to thank the Alexander-von-Humboldt foundation for providing a postdoctoral fellowship.

# Supplementary information

# Breaking the doping limit in silicon by deep level impurities


Mao Wang[1,2,*], A. Debernardi[3,**], Y. Berencén[1], R. Heller[1], Chi Xu[1,2], Ye Yuan[1,4], Yufang Xie[1,2],

R. Böttger[1], L. Rebohle[1], W. Skorupa[1], M. Helm[1,2], S. Prucnal[1] and Shengqiang Zhou[1,*]

[1]Helmholtz-Zentrum Dresden-Rossendorf, Institute of Ion Beam Physics and Materials Research, Bautzner Landstr. 400, 01328 Dresden, Germany

[2]Technische Universität Dresden, 01062 Dresden, Germany

[3]CNR-IMM, sede Agrate Brianza, via Olivetti 2, I-20864, Agrate Brianza, Italy

[4]Physical Science and Engineering Division (PSE), King Abdullah, University of Science and Technology (KAUST), Thuwal 23955-6900, Saudi Arabia

[*]Corresponding authors, email: m.wang@hzdr.de; s.zhou@hzdr.de

[**]Corresponding author [theory] e-mail: alberto.debernardi@mdm.imm.cnr.it


**A. Experimental details**

The single-side polished Si (100) substrates (*intrinsic*, $\rho \geq 10^4$ Ω·cm) were implanted with Te ions to nominal fluences from $1.1 \times 10^{15}$ cm$^{-2}$ to $1.1 \times 10^{16}$ cm$^{-2}$ at room temperature. The peak Te concentrations ($c_{pk}$) were firstly calculated using SRIM code [1], and then verified by Rutherford backscattering spectrometry/channeling (RBS/C) measurements and SIMNAR [1] simulation. The ion-implanted samples were exposed to single laser pulse with the energy density of 1.2 J/cm$^2$. The whole amorphous implanted region was molten and then recrystallized with a solidification speed in the order of 10 m/s while cooling down [2] during the annealing process. After that the Te-rich layer is recrystallized and spans approximately 120 nm from the surface [3].

The RBS/C measurements were carried out using a 1 mm diameter collimated 1.7 MeV He$^+$ beam of the Rossendorf van de Graff accelerator with a 10-20 nA beam current. Energy spectra of backscattered He were detected at an angle of 170° with respect to the incoming beam direction using silicon surface barrier detectors with an energy resolution of 15 keV.



Due to the so-called channeling effect, the backscattering probability sensitively depends on the direction of the incident beam with respect to the crystallographic orientation of the sample, and is significantly reduced around the crystal axes and planes [4]. Angular axial scans were performed using a two-axis goniometer were by recording integral backscattering yields in energy windows corresponding to scattering from Si and Te atoms - see FIG. S3) as a function of angle of incidence around different crystallographic axial directions (<100>, <110> and <111> ). In addition, two-dimensional angular backscattering patterns have been obtained.

The electrical properties of the Te-hyperdoped Si samples were examined using a commercial Lakeshore Hall System in van-der-Pauw-geometry [5]. Samples were measured in the temperature range from 2 to 300 K and a magnetic field perpendicular to the sample plane was swept from -6 T to 6 T. The gold electrodes were sputtered onto the four corners of the square-like samples. The native $SiO_2$ layer was removed by HF etching prior to the sputtering process. Silver glue was used to contact the wires to the gold electrodes.

**B. Computational details**

We performed structural and electronic calculations in the framework of the Density Functional Theory (DFT), solving the Kohn-Sham equations through the plane-wave pseudopotential approach, as implemented in the Quantum Espresso (QE) open-source code [6]. We used ultra-soft pseudopotentials [6-9] in the separable form introduced by Kleinmann and Bylander [10], generated with a Perdew-Burke-Ernzerhof (PBE) exchange-correlation functional and smearing techniques. We chose a 50 Ry cutoff radius for the electronic valence wave-function, and 500 Ry cutoff radius for the charge density.

The ab initio simulations of doped silicon were performed by using the super-cell method. In a similar way as done e.g. in Ref. [11], we varied the size of the supercell to simulate different concentrations of impurities (in our case the Te concentration ranges from ~0.4% to ~3.1%). Periodic (i.e. Born-von Karman) boundary conditions are used. So, in the hyperdoping regime considered in the present work, each impurity interacts with its own images simulating a uniform distribution of the same defect type. This approximation is expected to provide qualitative estimation of average properties –like the defect formation energy - of defect types representing a fraction of the total defect population, while it is expected to provide a reliable quantitative estimation of average properties of defect types representing the large majority of defects present in the sample.   In the case in which only one type of defect is present, or the concentration of the other types of defects is negligible the



present method would provide exact results within the density functional approximation technique used.

## C. Transport charaterations

The Te-hyperdoped Si samples with the single-side polished *intrinsic* Si substrate ($\rho > 10^4$ Ω·cm) were used for the electrical characterization. The sheet resistance ($R_S$) of samples over the temperature range 2–300 K was measured using the van der Pauw technique [5]. The bulk electrical resistivity $\rho$ was calculated using $\rho = R_S*d$ and the conductivity $\sigma = 1/\rho$. The effective layer depth $d$ is 125 nm, which is verified by the RBS results (see FIG. S1). Figure S1 shows the temperature-dependent conductivity with the Te peak concentration varied from $1.25 \times 10^{20}$ cm$^{-3}$ (0.25%) to $1.25 \times 10^{21}$ cm$^{-3}$ (2.5%) in a temperature range of 2-300 K. There, the room-temperature conductivity of Te-hyperdoped Si samples is higher than $10^2$ (S/cm), which confirms that the intrinsic Si substrate has no influence on the transport properties of the Te-hyperdoped Si layer considering the respective thickness. As shown in FIG. S1, a rigorous experimental evidence of an IMT lies in the measurement of nonzero conductivity as the temperature tends to zero (described previously [3]). Samples Te-0.25% and Te-0.50% show insulating behavior with a strong temperature-dependent conductivity. Samples with the Te peak concentration exceeding 1.0% present a dramatically different behavior, exhibiting a conductivity with negligible variation with temperature and a finite value of more than $10^2$ (S/cm) as the temperature tends to zero. At 2 K, the conductivity of the samples with 1.0% and 0.25% Te concentration differs by nearly 4 orders of magnitude in spite of the difference of just four times in the concentration. Samples with Te concentration exceeding 1.0% show a quasi-metallic behavior. Moreover, the room-temperature conductivity of Te-hyperdoped Si is around 600 (S/cm) while the conductivity of S- and Se-hyperdoped Si is about 110 (S/cm) and 300 (S/cm) respectively at similar doping levels (4.2~4.9×10$^{20}$ cm$^{-3}$) [5,12].



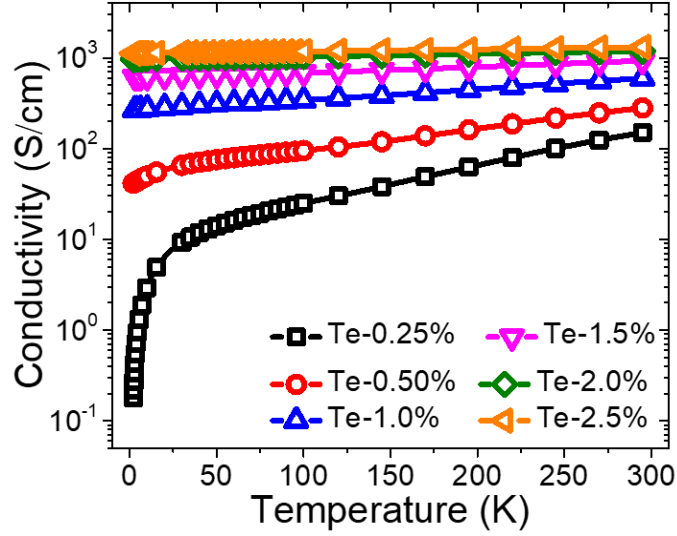

FIG. S1. The temperature-dependent conductivity of Te-hyperdoped Si in the temperature range of 2-300 K. Samples with lower Te concentrations (Te-0.25% and Te-0.50%) show a strong temperature-dependent conductivity, indicative of the insulating state. Samples with higher concentrations (Te-2.0% and Te-2.5%) exhibit a conductivity comparatively insensitive to temperature down to 2 K, indicating an impurity-mediated transition to the metallic-like state.

Based on the basic physical principle of the Hall effect, the carrier density of the samples can be estimated using standard techniques and a magnetic field B ranging from -4 T to 4 T. In some cases, it is convenient to use a layer or sheet density ($n_s$) instead of bulk density by using $n_s = \frac{IB}{e\gamma V_H}$, where $I$ is the current, $B$ is the magnetic field, $V_H$ is the Hall voltage, $e$ (1.602 x 10$^{-19}$ C) is the elementary charge and $\gamma$ is the Hall scattering factor that is generally assumed to be 1 heavily doped Si. If the conducting layer thickness $d$ is known, the bulk density ($N$) can be determined as $N=n_s/d$ [4,12]. Here, the carrier concentration ($n$) ranges from $2.0 \times 10^{19}$ to $8.3 \times 10^{20}$ cm$^{-3}$, which is calculated by assuming the effective thickness of the tellurium doped layer as 125 nm [3]. This relative large magnitude of the donor concentration is comparable to the shallow level impurity doped Si with just-metallic concentrations [5] and is consistent with the metallic-like samples in S and Se doped Si [11-12].

**D. Rutherford backscattering spectrometry/channeling (RBS/C) measurements**

Figure S2 shows the representative RBS/C spectra of a PLM-treated Te-hyperdoped Si layer with different Te concentrations, also including the virgin Si and the as-implanted wafers. The random spectrum of the as-implanted layer reveals a thickness of about 120 nm



for the Te profile, in an approximately Gaussian distribution. No channeling effect is observed in the implanted layer because of the amorphization caused by ion implantation. From the RBS-channeling signal, the PLM-treated layer shows a minimum backscattered yield $\chi_{min}$ of 4% (defined as the ratio of the aligned to the random yields near the sample surface from 740 keV to 920 keV), which is comparable to that of virgin Si substrate. This indicates a full re-crystallization and an epitaxial growth of the PLM-treated layers. Moreover, a channeling behavior of Te (from 1420 keV to 1520 keV) is observed in the PLM-treated samples, which proves the incorporation of Te into the Si lattice sites.

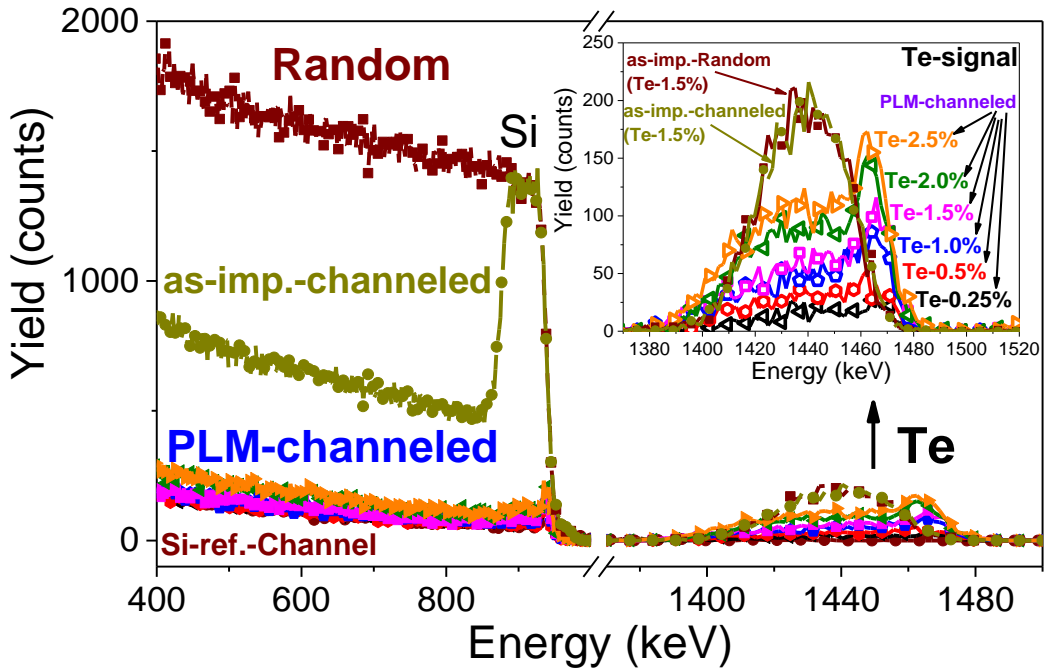

FIG. S2 A sequence of 1.7 MeV He RBS/channeling spectra of as-implanted and PLM-treated Te-hyperdoped Si layers with different Te concentrations. The inset shows the magnification of random and channeling Te signals for the as-implanted and PLM-treated layers.

### E. Lattice location of impurities

Figure S3(a) shows the schematic of the Si matrix with the implanted impurities which can take place at different lattice (a, b, c and d) sites. FIG. S3(b) shows the angular scans for the respective lattice sites: the dashed line corresponds to the impurity, while the solid one to Si [13]. If both angular scans match well with each other, impurities are located at the substitutional sites with high symmetry. The angular scan for impurities occupying interstitial sites is displayed in FIG. S3(b). Normally, two typical parameters were used to characterize the angular distributions, $\chi_{min}$ and $\psi_{1/2}$. In detail, $\chi_{min}$ corresponds to the minimum value of the



backscattered ion angular spectrum with respect to the random direction normalized to 1 [14]. Qualitatively, it presents the ability of dopants blocking the channels. The critical angle $\psi_{1/2}$ represents the half width of the angular scans at half maximum between $\chi_{min}$ and 1 (full width at half maximum of the angular scan spectrum), and it is related to the displacement vector of substitutional atoms [15].

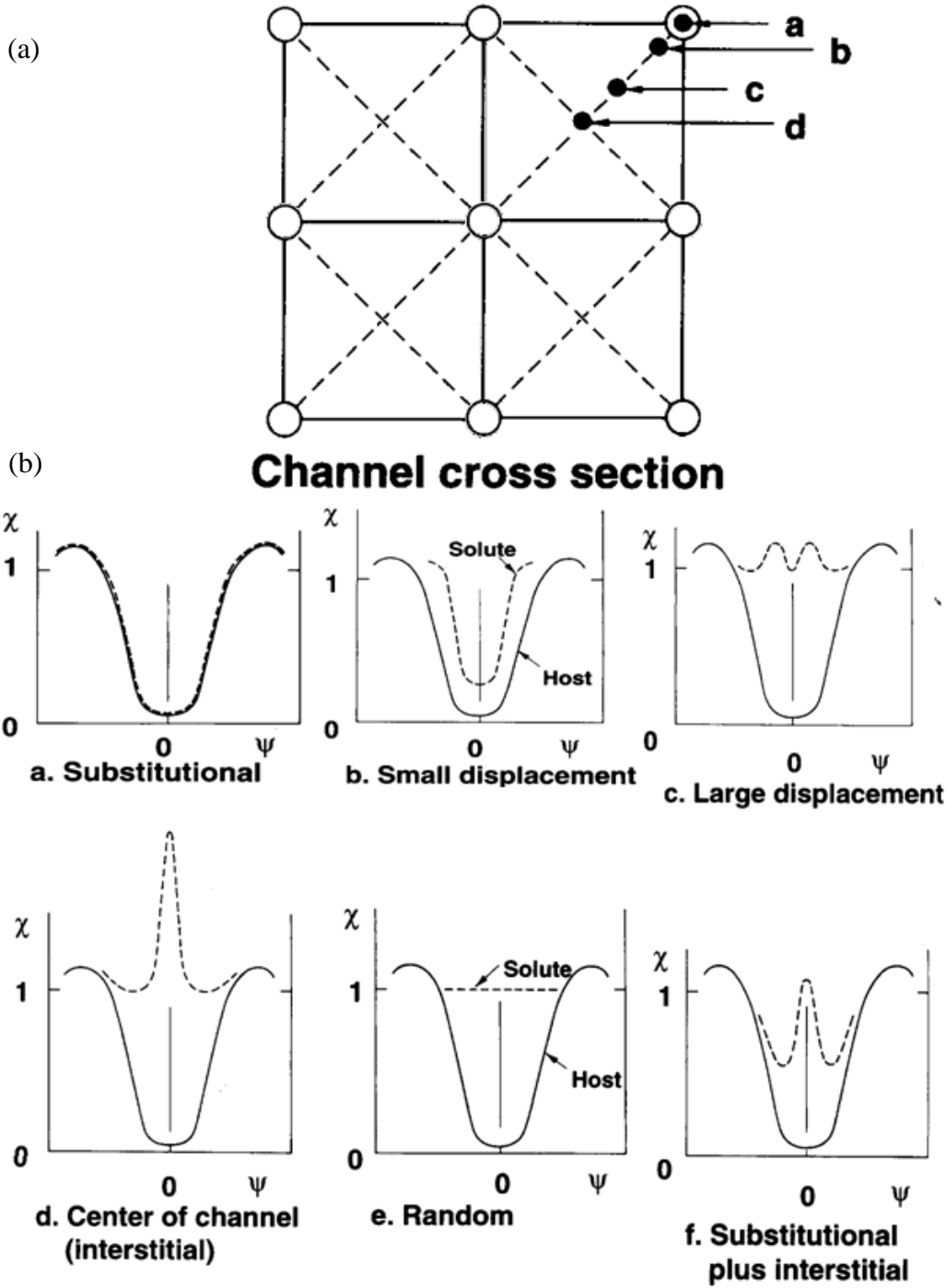

(a)

(b) Channel cross section

a. Substitutional
b. Small displacement
c. Large displacement
d. Center of channel (interstitial)
e. Random
f. Substitutional plus interstitial



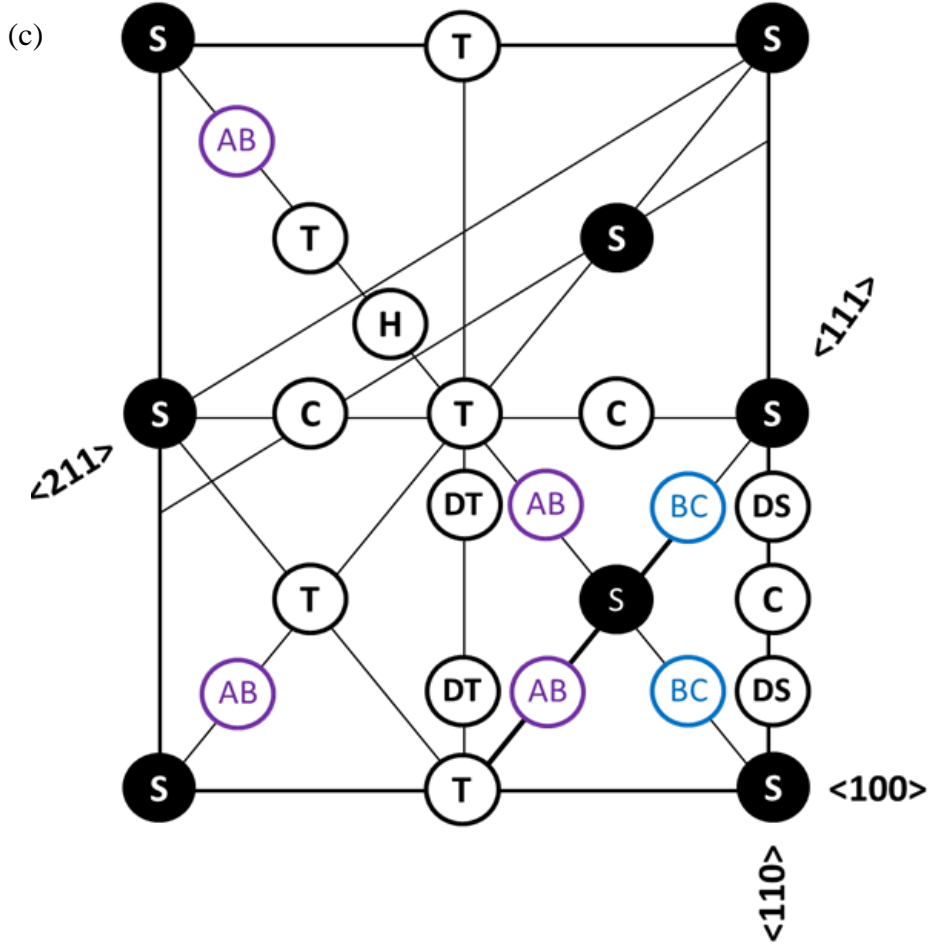

FIG. S3. The schematic structure of the Si matrix (a): The solid black circle represents the impurity atom and the empty black circles stand for the host material atoms. The RBS/channeling angular scans for an impurity in the Si host material, substitutional site and interstitial site (channel center) are shown in (b) [1]. (c) Positions of major sites in the Si lattice (hexagonal (H), tetrahedral (T), bond-centered (BC), anti-bonding (AB), split (SP), DS and DT sites (sites displaced from these high symmetry sites)), shown in the {110} plane [16].

Figure S4 shows the angular distributions for Te and Si at the same depth in the vicinity of the three main crystallographic directions, namely <100>, <110> and <111>, recorded off a major crystallographic plane. Two typical parameters $\chi_{min}$ and $\psi_{1/2}$ have been simultaneously calculated for Si and Te at the same depth. The three axial angular scans for Te follow those of the Si host in all the Te-hyperdoped Si samples, indicating that the dominant fraction of the Te atoms are located at substitutional [14,16] sites. For samples Te-0.25% and Te-0.50%, the $\chi_{min}$-*Te* is significantly higher than that of Si. This suggests that a non-neglectable fraction of the Te atoms does not occupy substitutional sites and those atoms are located at the low symmetric interstitial sites lying around the axes of the crystal lattice. As depicted in FIG. S4(a), (b), (g), (h), (m) and (n), the interstitial site impurities are clearly evidenced. While for samples with higher doping concentrations, a puzzling and interesting behavior has been observed from the angular scans. The fraction of the interstitial Te atoms tends to vanish. As



shown in FIG. S4 and contrary to the general expectation, the interstitial fraction decreases with increasing impurity concentration in Te-hyperdoped Si samples, which shows the similarity in the channeling direction of <100>, <110> and <111>. The $\psi_{1/2}$-Te is narrower than that of $\psi_{1/2}$-Si (see FIG. S4) for all the Te-hyperdoped Si samples in the three axial angular scans meaning that the Te dopants do not perfectly substitute Si atoms from the host. There exists a displacement ($r_0$) from the ideal *S* site (i.e. *DS* sites) [17].



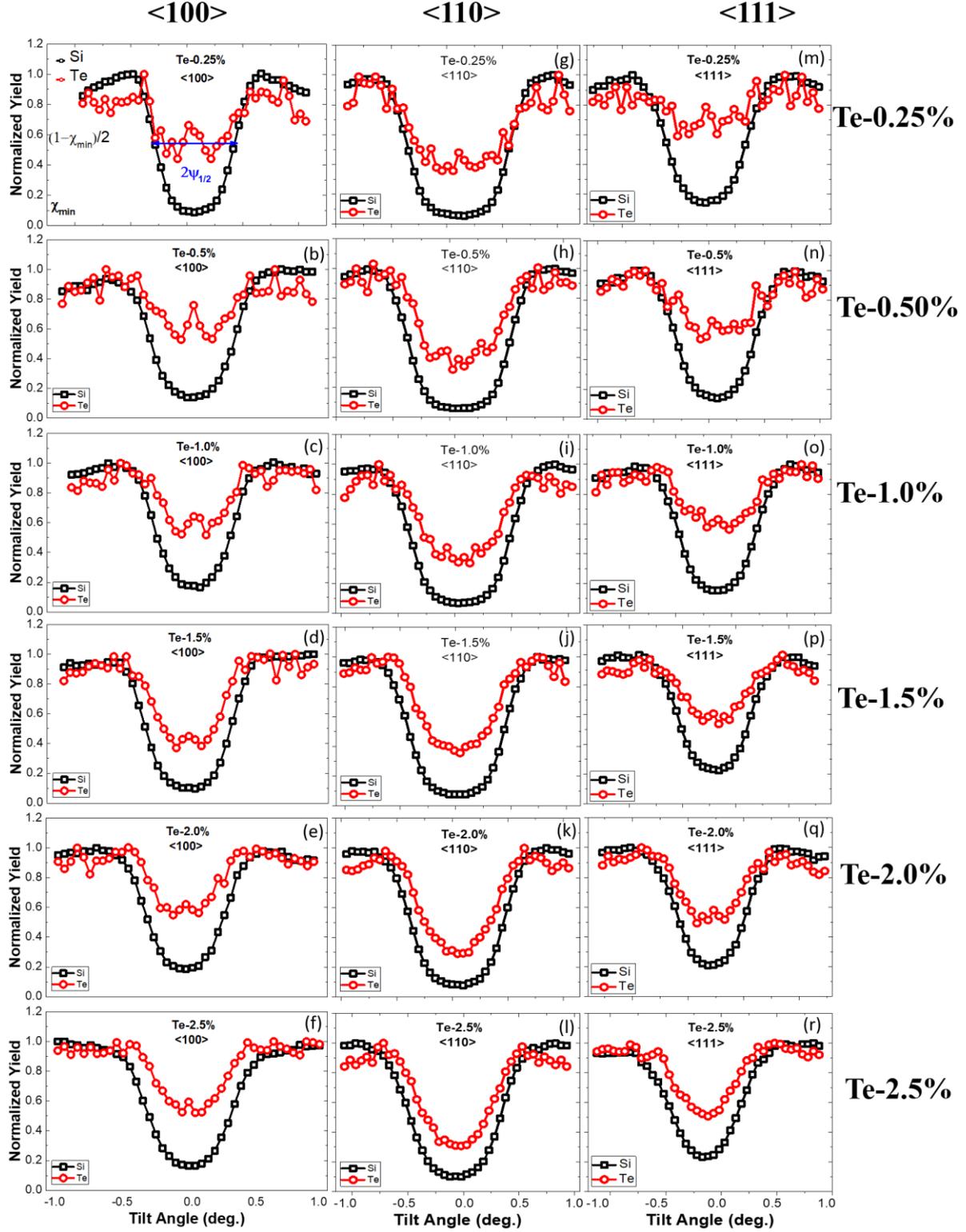

FIG. S4. (a)-(r) The angular scans about the <100> ((a)-(f)), <110> ((g)-(l)) and <111> ((m)-(r)) axes for all the Te-hyperdoped Si samples. Angular distributions have been normalized in a random direction. The red circle corresponds to Te atoms and black square to Si, smooth curves are drawn through the symbols.

## F. Displacement estimation



The RBS/C angular scans shown in FIG. S4 for Si and Te suggest that the Te atoms may be displaced from the ideal substitutional Si lattice sites. To obtain quantitative information about the Te location, the angular distributions as a function of the equilibrium displacement of an atom from a lattice row [18] was calculated. The parameter $\psi_{1/2}$ can be expressed as below for displacement calculation [13]:

$$\psi_{1/2} = \sqrt{\left[\frac{1}{2}ln\left(\frac{C^2 a^2}{r_{min}^2} + 1\right)\right]} \times \sqrt{\frac{2Z_1 Z_2 e^2}{Ed}} \quad (1),$$

Where constant $C^2$ is around 3, $Z_1$ and $Z_2$ are the atomic numbers of the incident particle and of the considered atom, respectively, $e^2$ is equal to 14.4 $eV\ Å$, $E = 1.7\ MeV$ is the incident ion energy, and $d$ is the average atomic distance in the considered atomic row ($d_{aver}[100] = 5.43\ Å\ \ d_{aver}\langle 110\rangle = 3.84\ Å\ \ d_{aver}\langle 111\rangle = 4.70\ Å$) [14].

$$r_{min}^2 = \rho^2 ln2 + r_0^2 \quad (2),$$

Here $\rho$ is the thermal vibration amplitude, and $r_0$ is the projection (on a plane perpendicular to the channeling direction) of the equilibrium displacement of the Te atoms with respect to the lattice sites. $a$ is the screening distance, which is related to the Bohr radius ($a_0$=0.528 $Å$).

$$a = 0.8853 \times a_0 \times \frac{1}{\sqrt[3]{\left(z_1^{\frac{1}{2}}+z_2^{\frac{1}{2}}\right)^2}} = 0.11114 \quad (3),$$

Here, the range for the average error of the projection displacement was determined by the dispersion of the parameter $\psi_{1/2}$.

For comparison, we computed the displacements obtained in our ab initio simulations. To minimize the numerical error, that in our simulation is comparable to the variation of the displacement as a function of the Te concentration, we averaged the values of displacements of Te impurities obtained from first-principles simulations for $S_{Te}$-$S_{Te}$ over the whole range of the investigated concentration. The average results, corresponding to the average Te concentration of 1.36% are displayed in Table I (bottom panel) in good agreement with the experimental data for 1.0% and 1.5% Te concentrations.

TABLE I. Displacement obtained from formula (1), which corresponds to the lattice sites projected perpendicularly to the channeling direction. The [100] direction is perpendicular to the Si substrate.

| | $r_0$ (Experimental values) | | |
|---|---|---|---|
| | $r_{<100>}$ | $r_{<110>}$ | $r_{<111>}$ |
| Te-0.25% | | | |
| Te-0.50% | 0.36±0.03 | 0.40±0.03 | 0.34±0.03 |
| Te-1.0% | 0.37±0.03 | 0.40±0.03 | 0.36±0.03 |
| Te-1.5% | 0.42±0.03 | 0.40±0.03 | 0.38±0.03 |
| Te-2.0% | 0.43±0.03 | 0.41±0.03 | 0.45±0.03 |
| Te-2.5% | 0.46±0.03 | 0.42±0.03 | 0.46±0.03 |
| | $r_0$ (Computational values) (Average displacement=0.46 Å) | | |
| | $r_{<100>}$ | $r_{<110>}$ | $r_{<111>}$ |
| Te-1.36% | 0.38 | 0.37 | 0.33 |